\documentclass[%
reprint,
amsmath,amssymb,
aps,
pre,
]{revtex4-1}

\usepackage{graphicx}
\usepackage{dcolumn}
\usepackage{bm}
\usepackage{natbib}
\usepackage{mathtools}
\usepackage{physics}
\usepackage{bbm}

\newcommand{\Cov}{\mathrm{Cov}}
\newcommand{\Var}{\mathrm{Var}}
\DeclareMathOperator{\arcsinh}{arcsinh}

\begin{document}
\title{Achievability of thermodynamic uncertainty relations}
\author{Jiawei Yan}
\email[]{jiaweiyan@g.harvard.edu}
\affiliation{Department of Systems Biology, Harvard University, 200 Longwood Avenue, Boston, Massachusetts 02115, USA}

\date{\today}

\begin{abstract}
    The thermodynamic uncertainty relations provide a universal trade-offs between entropy dissipation rate and fluctuations in transport current. This relation has been mostly used to estimate a minimum entropy dissipation rate by experimentally measuring current fluctuations. Here we consider joint currents and show that such uncertainty relations cannot be simultaneously achievable for any two currents, which leads to a potential underestimation of the minimum entropy dissipation rate.
\end{abstract}
\maketitle

Recent advances in stochastic thermodynamics have revealed a remarkable trade-off---the thermodynamic uncertainty relations (TURs)---between the fluctuation of transport currents and system's entropy production rate \cite{barato2015thermodynamic, gingrich2016dissipation, Gingrich2017inferring, dechant2018multidimensional, hasegawa2019uncertainty}. In general, it states that in any small systems such as molecular motors where fluctuations inevitably affect performances, the more precise a steady current is, the more energy dissipation it requires. This relation was first proven for Markov jump processes in long time limit by a large deviation approach \cite{gingrich2016dissipation}, and has been almost immediately extended to many different directions by similar techniques, including diffusion processes, finite-time processes, discrete-time processes, periodic driving systems, quantum systems, first-passage-time fluctuations, etc \cite{Gingrich2017FPT, horowitz2017proof, Gingrich2017inferring, proesmans2017discrete, dechant2018multidimensional, chiuchiu2018mapping, barato2018bounds, agarwalla2018assessing, Brandner2018thermo}. This collection together not only offers constraints on the fluctuations of currents in non-equilibrium systems, but provides a new method of inferring entropy production rate---which usually cannot be measured directly by experiments---by experimentally measuring the current fluctuations \cite{li2018quantifying, martinez2019inferring, manikandan2019inferring}. However, it remains elusive whether or when these TURs are tight, i.e., the minimum is achievable by real systems far from equilibrium \cite{Gingrich2017inferring}.

Recently, the original scalar-valued version of long time limit fluctuations has been extended to multidimensions by an information-theoretic approach \cite{dechant2018multidimensional}. Here we provide an alternative and more general proof using large deviation theory. Our proof could be seen as a natural generalization of the first proof for scalar-valued relation to include joint fluctuations of currents \cite{gingrich2016dissipation}. Thus all its extensions, such as the bound on first-passage-time and finite-time fluctuations, can be generalized to vector-value immediately in the same manner. We then use this multidimensional relation to prove a striking fact that any two scalar-valued TURs for different observed currents cannot be achieved simultaneously, i.e., their minima cannot be reached in the same time. In other words, there is at most only one current among all the possible ones in the system which fluctuation is able to achieved the minimum predicted by TURs. Therefore, if the observed current is not carefully chosen, there will be an underestimation of the minimum entropy production rate inferred from the scalar-valued TURs.

\section{Derivation of the multidimensional TURs by large deviation}
We first derive the multidimensional TURs by large deviation theory. Our deviation follows the similar logic with \cite{gingrich2016dissipation}, which is based on Markov jump process but can also be applied for diffusion processes. We first review the original derivation since it is also important to derive the new result. Here without loss of generality, we consider a classical thermodynamic system with $N$ mesoscopic states, with transition between states $y$ and $z$ are modelled as a continuous-time Markov jump process with rate $r(y,z)$. Let the total number of edges $(y,z)_{y<z}$ with non-zero transition rates be $E \leqslant N(N-1)/2$. Thermodynamic consistency requires local detailed balance condition, i.e., $r(y,z)\neq0\Leftrightarrow r(z,y)\neq0$, and we assume the system is irreducible which further ensures that there exists a unique stationary probability distribution $\pi(y)$ \cite{van1992stochastic}. For systems poised at thermo-equilibrium, for any two mesoscopic states $y$ and $z$, the empirical current $j(y,z)$---which empirically counting the number of net jumps from $y$ to $z$ in a unit time---must converges to 0 in the long time limit because of the detailed balance. For systems far from equilibrium, instead, a non-vanishing current $j(y,z)$ may exist \cite{qian2007phosphorylation}. In the long time limit, such current will converges to the associated stationary current $j^\pi(y,z) = \pi(y)r(y,z) - \pi(z)r(z,y)$, but fluctuations always exist, which is the observable we are interested in. 

We start from level 2.5 large deviation rate function for Markov jump processes. Level 2.5 large deviation theory focuses on the joint probability distribution $\mathbb{P}(p,j)$ of the empirical density $\pmb{p}\in\mathbb{R}^N$ and empirical current $\pmb{j}\in\mathbb{R}^E$. In the long but finite time $T$, it has an asymptotic relation $\mathbb{P}(\pmb{p},\pmb{j})\asymp e^{-T I(\pmb{p},\pmb{j})}$, where the rate function $I(\pmb{p},\pmb{j})$, according to \cite{bertini2015large}, is equal to 
\begin{equation}
    I(\pmb{p},\pmb{j}) = \sum_{y<z} \Psi(j(y,z),\bar{j}(y,z),a(y,z)),
\end{equation}
and
\begin{align}
   \Psi(j,\bar{j},a) = \ & j\left(\arcsinh\frac{j}{a} - \arcsinh\frac{\bar{j}}{a}\right) \nonumber\\
    &  - \left(\sqrt{j^2+a^2} - \sqrt{\bar{j}^2+a^2}\right),
\end{align}
where $\bar{j} = p(y)r(y,z) - p(z)r(z,y)$ is the average current associated with a given empirical density and $a(y,z) = 2\sqrt{p(y)r(y,z)p(z)r(z,y)}$. Then we use an upper bound $\Psi_{\text{LR}}(j)\geqslant\Psi(j,\bar{j},a)$ of each edge $(y, z)$, derived from \cite{gingrich2016dissipation}:
\begin{equation}
\label{EQ: single quadratic bound}
    \Psi_{\text{LR}}(j(y,z)) = \frac{\left(j(y,z) - j^\pi(y,z) \right)^2}{4j^\pi(y,z)^2}\sigma^\pi(y,z),
\end{equation}
where $\sigma^\pi(y,z) = j^\pi(y,z)\ln\frac{\pi(y)r(y,z)}{\pi(z)r(z,y)}$ is the stationary entropy production rate of edge $(y,z)$ \cite{ge2010physical, esposito2010three}. By the contraction principle \cite{ellis1985grundlehren}, we then obtain an upper bound for the rate function of the current:
\begin{equation}
    I(\pmb{j})\leqslant \sum_{y<z} \Psi_{\text{LR}}\left(j(y,z)\right).
\end{equation}
  
Since experimentally measuring the current between only mesoscopic states may often be challenging, what we are more interested in is the generalized current, defined as a linear combination of the current of each edge:
\begin{equation}
\label{EQ: sGC}
    \jmath = \sum_{y<z} d(y,z) j(y,z).
\end{equation}
If only one generalized current is of interested, then the rate function of this generalized current can be bounded by letting $\pmb{j}^* = \pmb{j}^\pi \jmath/\jmath^\pi$, where $\pmb{j}^\pi\in \mathbb{R}^E$ is the vector of all $j^\pi(y,z)$. $\pmb{j}^*$ here automatically satisfies eq.~(\ref{EQ: sGC}), therefore by plugging in $\pmb{j}^*$ into eq.~(\ref{EQ: single quadratic bound}) we obtain the upper bound of the rate function of $\jmath$:
\begin{equation}
\label{EQ: GC quadratic bound}
    I(\jmath) \leqslant \frac{1}{4}\left(\frac{\jmath}{\jmath^\pi} - 1\right)^2 \sum_{y<z}\sigma^\pi(y,z).
\end{equation}

Let $\Sigma^\pi_{\text{tot}} = \sum_{y<z}\sigma^\pi(y,z)$ which is equal to the total entropy production rate \cite{esposito2010three}. The relative uncertainty of $\jmath$ is defined as variance normalized by mean square:
\begin{equation}
    \epsilon^2 = \Var[\jmath]/(\jmath^\pi)^2,
\end{equation}
then the bound eq.~(\ref{EQ: GC quadratic bound}) becomes a bound on variance by the fact $I(\jmath^\pi)'' = 1/\Var[\jmath]$ \cite{ellis1985grundlehren}, which leads to the scalar-value thermodynamic uncertainty relation first conjectured by A.C.~Barato and U.~Seifert \cite{barato2015thermodynamic}, and later proven by T.R.~Gingrich \textit{et al.} \cite{gingrich2016dissipation}:
\begin{equation}
\label{EQ: sTUR}
    \epsilon^2 \Sigma^\pi_{\text{tot}} \geqslant 2.
\end{equation}

Now we generalized the results above to situations when multiple generalized currents are of insterest and measured simultaneously. Consider a vector of $M$ generalized currents $\pmb{\jmath}\in\mathbb{R}^M$, and let $\jmath_\ell = \sum_{y<z} d_\ell(y,z)j(y,z)$ be the $\ell$-th one. Without loss of generality, we only consider linear independent combination, i.e., $d_i(y,z)$ cannot be expressed as a linear combination of the rest $d_\ell(y,z)$, which further constrains that $M$ must be less or equal to $E$. From these conditions there always exist a number of $E$ vectors $\pmb{g}^{(y,z)}\in\mathbb{R}^M$ such that
\begin{equation}
    j^*(y,z) = \sum_{\ell = 1}^M g^{(y,z)}_{\ell}\jmath_\ell
\end{equation}
(such $\pmb{g}^{(y,z)}$ is the Moore-Penrose inverse, which is not unique unless $M = E$). Plug in such $\pmb{j}^*$ into eq.~(\ref{EQ: single quadratic bound}) and yield:
\begin{equation}
\label{EQ: mGC bound}
    I(\pmb{\jmath}) \leqslant \sum_{y<z} \left(\sum_{\ell=1}^M g^{(y,z)}_{\ell}\jmath_\ell - j^\pi(y,z) \right)^2\frac{\sigma^\pi(y,z)}{4\left[j^\pi(y,z)\right]^2}.
\end{equation}

Let the vector of $M$ generalized current at the stationary state be $\pmb{\jmath}^\pi$. The covariance matrix $\Xi$ of $\pmb{\jmath}$ defined as
\begin{equation}
    \Xi_{pq} = \mathbb{E}[(\jmath_p - \jmath_p^\pi)(\jmath_q - \jmath_q^\pi)],\quad \text{for all } 1\leqslant p,q\leqslant M \nonumber
\end{equation}
can be obtained from the Hessian matrix of $I(\pmb{\jmath})$ at $\pmb{\jmath}^\pi$ \cite{ellis1985grundlehren}:
\begin{equation}
\label{EQ: Hessian Cov relation}
    \mathrm{H}[I(\pmb{\jmath})|\pmb{\jmath}^\pi] = \Xi^{-1},
\end{equation}
where $\mathrm{H}[I(\pmb{\jmath})|\pmb{\jmath}^\pi]$ is the Hessian matrix of $I(\pmb{\jmath})$ evaluated at the stationary state. Combing eq.~(\ref{EQ: Hessian Cov relation}) and eq.~(\ref{EQ: mGC bound}), we can obtain a matrix inequality\footnote{For Hermitian $A$ and $B$,  $A\preceq B$ means $(B-A)$ is positive semi-definite.}:
\begin{equation}
\label{EQ: matrix TUR}
    \Xi^{-1} \preceq G D G^\top,
\end{equation}
where $D$ is an $E\times E$ diagonal matrix with diagonal entries $\sigma^\pi(y,z) / [2j^\pi(y,z)^2]$, and $G$ is an $M\times E$ matrix with columns $\pmb{g}^{(y,z)}$. Eq.~(\ref{EQ: matrix TUR}) is hard to interpret since $G$ is not uniquely chosen, therefore we multiply $(\pmb{\jmath}^\pi)^\top$ and $\pmb{\jmath}^\pi$ from left and right, respectively:
\begin{align}
    (\pmb{\jmath}^\pi)^\top[\Xi(\pmb{\jmath})]^{-1}(\pmb{\jmath}^\pi) \leqslant [G^\top(\pmb{\jmath}^\pi)]^\top D[G^\top(\pmb{\jmath}^\pi)] = (\pmb{j}^\pi)^\top D \pmb{j}^\pi. \nonumber
\end{align}
We complete the derivation by noting that $(\pmb{j}^\pi)^\top D \pmb{j}^\pi = \frac{1}{2}\sum_{y<z}\sigma^\pi(y,z) = \frac{1}{2}\Sigma^\pi_{\text{tot}}$:
\begin{equation}
    \label{EQ: mTUR dechant}
    (\pmb{\jmath}^\pi)^\top[\Xi(\pmb{\jmath})]^{-1}\pmb{\jmath}^\pi \leqslant \frac{1}{2}\Sigma^\pi_{\text{tot}}.
\end{equation}

We finally end up with the same result as Dechant has shown in the eq.~(29) in \cite{dechant2018multidimensional} for Langevin systems. Here our approach of large deviation has been shown to also apply to continuous diffusion processes \cite{Gingrich2017inferring}, providing a proof which is a natural generalization of the original work on scalar-valued relations.

\section{Achievability of the thermodynamic uncertainty relations}
The thermodynamic uncertainty relations, both scalar-valued and vector-valued, could be potentially useful to estimate a minimum entropy production rate by experimentally measuring the fluctuation of currents \cite{li2018quantifying, martinez2019inferring, manikandan2019inferring}. However, such estimation is mostly meaningful if the eq.~(\ref{EQ: sTUR}) or (\ref{EQ: mTUR dechant}) are tight bounds, i.e., their minima are achievable. It has been proven that eq.~(\ref{EQ: sTUR}) is indeed tightest and can be achieved within linear-response region when the generalized current is the empirical entropy production itself \cite{gingrich2016dissipation, Gingrich2017inferring}, which is to choose the linear combination $d(y,z)$ to be the corresponding thermodynamic force:
\begin{equation}
    d(y,z) = \ln\frac{\pi(y)r(y,z)}{\pi(z)r(z,y)},
\end{equation}
and the empirical entropy production rate is:
\begin{equation}
    \Sigma_{\text{tot}} = \sum_{y<z} j(y,z)\ln\frac{\pi(y)r(y,z)}{\pi(z)r(z,y)}.
\end{equation}

However, the tightness, or achievability, is still unclear for other generalized currents. Here we show that for any two generalized currents, the scalar-valued TURs cannot be achieved simultaneously.

First note that eq.~(\ref{EQ: mTUR dechant}) can be written as
\begin{equation}
\label{EQ: mTUR dechant rho}
    (\pmb{\epsilon}^{-1})^\top \varrho^{-1} \pmb{\epsilon}^{-1} \leqslant \frac{1}{2}\Sigma^\pi_{\text{tot}},
\end{equation}
where $\pmb{\epsilon}^{-1}$ is an $M$-dimensional vector:
\begin{equation}
    \pmb{\epsilon}^{-1} = \left[\frac{1}{\epsilon_1}, \frac{1}{\epsilon_2}, \cdots \frac{1}{\epsilon_M}\right]^\top,
\end{equation}
and 
\begin{equation}
    \epsilon_\ell = \sqrt{\frac{\Var[\jmath_\ell]}{(\jmath_\ell^\pi)^2}}
\end{equation}
is the square root of the relative uncertainty of $\ell$-th generalized current, and $\varrho$ is the correlation coefficient matrix of $\{\jmath_1, \jmath_2,\cdots,\jmath_M\}$.

Now consider we are interested in two generalized current $\jmath_1$ and $\jmath_2$, the scalar-valued relation, eq.~(\ref{EQ: sTUR}), says each uncertainty or fluctuations $\epsilon_i$ is bounded by two over total entropy dissipation rate. Instead, the multidimensional relation eq.~(\ref{EQ: mTUR dechant rho}) states:
\begin{equation}
\label{EQ: 2d mTUR}
     \begin{bmatrix}1/\epsilon_1 & 1/\epsilon_2\end{bmatrix}\begin{bmatrix}1 & \rho \\ \rho & 1\end{bmatrix}^{-1} \begin{bmatrix}1/\epsilon_1 \\ 1/\epsilon_2\end{bmatrix}\leqslant\frac{1}{2}\Sigma^\pi_{\text{tot}},
\end{equation}
where $\rho$ is the correlation coefficient between $\jmath_1$ and $\jmath_2$. Divided by $\sqrt{\Sigma^\pi_\text{tot}}$ on both sides, it then gives a lower bound:
\begin{equation}
    \label{EQ: necessary tightness}
    \frac{2}{\epsilon_1^2\Sigma^\pi_{\text{tot}}} - 2\rho \sqrt{\frac{2}{\epsilon_1^2\Sigma^\pi_{\text{tot}}}}\sqrt{\frac{2}{\epsilon_2^2\Sigma^\pi_{\text{tot}}}} + \frac{2}{\epsilon_2^2\Sigma^\pi_{\text{tot}}} \leqslant (1-\rho^2).
\end{equation}

Comments are in order. First, eq.~(\ref{EQ: 2d mTUR}) and (\ref{EQ: necessary tightness}) imply that to let the scalar-valued relation achieves the minima for both generalized currents: $\epsilon_1^2\Sigma^\pi_{\text{tot}} =2$ and $\epsilon_2^2\Sigma^\pi_{\text{tot}} =2$, a necessary condition is $\rho= 1$. It means that unless the two generalized currents measured are perfectly correlated, the scalar-valued TURs cannot be achieved simultaneously. 

Then one crucial question may be when the generalized current correlates positively or perfectly with another. Let the two generalized currents $\jmath_1 = \pmb{a}^\top\pmb{j}$ and $\jmath_2 = \pmb{b}^\top\pmb{j}$, their covariance is then:
\begin{equation}
\label{EQ: improved sTUR}
    \Cov(\jmath_1, \jmath_2) = \pmb{a}^\top \Xi(\pmb{j})\pmb{b}
\end{equation}
where $\Xi(\pmb{j})$ is the covariance matrix of current of each edge. 
By Cauchy-Bunyakovsky-Schwarz inequality, $\Cov(\jmath_1, \jmath_2)$ reaches its maximum (perfectly correlated) when and only when $\pmb{a}$ and $\pmb{b}$ are linear dependent. Since here are only two vectors, it means $\pmb{a}=\pmb{b}$. On the other words, the two scalar-valued TURs for $\jmath_1$ and $\jmath_2$ only become simultaneously achievable when $\jmath_1$ and $\jmath_2 $ are equivalent, which trivially degenerates to scalar-valued case. This implies for a given system, there is at most only only generalized current which can achieve the minimum of the scalar relation eq.~(\ref{EQ: sTUR}). For all other generalized currents, the multidimensional TURs may serve as a tighter bound than the scalar version.

More importantly, let $\Delta_i = (\epsilon_\jmath^2\Sigma^\pi_{\text{tot}}/2 - 1) \geqslant 0$ be the ``tightness'', then from eq.~(\ref{EQ: necessary tightness}) we have
\begin{equation}
    \Delta_1 \Delta_2 \geqslant (1 - \rho \sqrt{\Delta_1+1}\sqrt{\Delta_2+1})^2,
\end{equation}
which gives another trade-offs between the ``tightness'' of two scalar-valued TURs unless the two generalized currents are carefully chosen such that their correlation $\rho = 1/\sqrt{(\Delta_1+1)(\Delta_2+1)}$.

Finally we want to emphasize that for systems near equilibrium, linear-response theory states that the variance of entropy production rate $\Var[\Sigma_{\text{tot}}]$ indeed reaches its minimum $2\Sigma_{\text{tot}}^\pi$ predicted by TURs \cite{Gingrich2017inferring, marconi2008fluctuation}, therefore for any other generalized current, its fluctuation cannot achieve the minimum $2/\Sigma_{\text{tot}}^\pi$ predicted by the scalar-valued TURs, which brings an underestimation of the minimum entropy production rate if the entropy production rate cannot be directly measured. For systems far from equilibrium, since the fluctuations of entropy production can probably be significant, the scalar-valued TURs of other generalized currents may become tight. However, for all generalized currents, there is still at most only one of them whose fluctuation can achieve the minimum $2/\Sigma_{\text{tot}}^\pi$.

\section{Conclusions}
To conclude, we use the large deviation theory to derive the multidimensional uncertainty relations which were first proven by Dechant \cite{dechant2018multidimensional} by a different method, Out proof comes from a natural generalization of the approach for the scalar-value uncertainty relations. Thus our derivation can be immediately generalized to finite-time fluctuations, first-passage-time, and other scenarios which similar scalar-valued relations are all derived by large deviation. Remarkably, with our results, the whole family of TURs \cite{Gingrich2017inferring, Gingrich2017FPT, horowitz2017proof} can be all derived from one formalism---the large deviation theory. We speculate these work together may formulate a uniform theory of nonequilibrium fluctuations. By applying the multidimensional relations, we further demonstrate that the scalar-valued relation is tight for only one generalized current among all the possible currents in the system. Hence the multidimensional TURs may be a better candidate to infer the minimum entropy production rate experimentally.

\begin{acknowledgments}
J.~Y.~gratefully acknowledge T.~Gingrich for helpful conversations. This research is supported from the Harvard Quantitative Biology Initiative and NSF-Simons Center for Mathematical and Statistical Analysis of Biology at Harvard.
\end{acknowledgments}

\bibliography{PRE_format}

\end{document}